\documentclass[structabstract]{aa}

\usepackage{graphicx}
\usepackage[usenames]{color}

\begin{document}

\title{
The near-infrared companion to HD\,94660 (=KQ\,Vel)
}

\author{
M.~Sch\"oller\inst{1}
	\and
C.~A.~Hummel\inst{1}
	\and
S.~Hubrig\inst{2}
	\and
D.~W.~Kurtz\inst{3,4}
	\and
G.~Mathys\inst{5}
	\and
J.~Robrade\inst{6}
	\and
S.~P.~J\"arvinen\inst{2}
}

\institute{
European Southern Observatory,
Karl-Schwarzschild-Str.~2,
85748~Garching, Germany\\
\email{mschoell@eso.org}
	\and
Leibniz-Institut f\"ur Astrophysik Potsdam (AIP),
An der Sternwarte~16,
14482~Potsdam, Germany
	\and
Centre for Space Research,
Physics Department, North West University,
Mahikeng~2745, South Africa
	\and
Jeremiah Horrocks Institute,
University of Central Lancashire,
Preston~PR1~2HE, UK
	\and
European Southern Observatory,
Alonso~de~Cordova~3107,
Vitacura, Santiago, Chile
	\and
Hamburger Sternwarte, Gojenbergsweg~112, 21029~Hamburg, Germany
             }

   \date{Received September 15, 1996; accepted March 16, 1997}

  \abstract
   {
The Bp star HD\,94660 is a single-lined spectroscopic binary.
Some authors have suggested that the unseen companion of at least $2\,{\rm M}_\odot$ may be a compact object.
   }
   {
We intend to study this multiple system in detail, especially to learn more about the so-far unseen companion.
   }
   {
We have collected and analyzed PIONIER H-band data from the Very Large Telescope Interferometer,
TESS visible photometric data, and
X-ray observations with Chandra of HD\,94660.
   }
   {
Using PIONIER, we were able to detect the companion to HD\,94660,
which is absent from high quality spectra at visible wavelengths,
with a magnitude difference of 1.8 in the H band at a separation of 18.72\,mas.
The TESS light curve shows variations with a period of 2.1\,d and also flaring.
The Chandra spectrum is well described by emission from hot thermal plasma,
yet might include a non-thermal component.
The X-ray properties are compatible with a magnetically active companion,
while some magnetospheric contribution from the primary is also possible.
   }
   {
We can rule out that the companion to HD\,94660 is a compact source.
It is also very unlikely that this companion is a single star,
as the estimated mass of more than $2\,{\rm M}_\odot$,
the magnitude difference of 1.8 in the H band,
and its non-detection in visible spectra are difficult to realize in a single object.
One alternative could be a pair of late F stars,
which would also be responsible for the detected photometric variations.
Interferometric observations over the full binary orbit are necessary to
determine the real mass of the companion and to add constraints on the overall geometry of the system.
   }


   \maketitle

\section{Introduction}
\label{sect:intro}

HD\,94660 (= KQ\,Vel) is a rather bright ($m_V=6.11$) chemically peculiar Bp star
(Renson \& Manfroid (\cite{RensonManfroid2009}) list a spectral type of A0p EuCrSi)
with a strong magnetic field,
which Bailey et al.\ (\cite{Bailey2015}) characterize
with a magnetic dipole strength $B_d = 7500$\,G with additional quadrupole and octupole moments,
an inclination angle of the stellar rotation axis $i = 16^{\circ}$,
and a magnetic obliquity angle $\beta = 30^{\circ}$.
As its measured mean longitudinal magnetic field of the order of $-2$\,kG is almost constant,
HD\,94660 is frequently used as a magnetic standard star.
It also shows spectral lines resolved into their magnetically split components.

Globally ordered magnetic fields are observed in roughly 10\%
of the intermediate and massive main-sequence stars with spectral types
between approximately B2 and F0.
These stars, generally called the chemically peculiar Ap and Bp stars (or Ap/Bp stars),
exhibit strong overabundances of certain elements, such as iron peak elements and rare earths,
and underabundances of He, C, and O, relative to solar abundances.
Massive Bp stars usually show overabundances of He and Si.
Ap/Bp stars show strictly periodic light, spectral, and magnetic variations with the
rotation period, which are well understood in terms of the oblique rotator model, in which the
magnetic axis is inclined with respect to the rotation axis (Stibbs \cite{Stibbs1950}).

HD\,94660 has an effective temperature $T_{\rm eff} = 11\,300 \pm 400$\,K,
a mass of $3.0\pm0.2\,{\rm M}_\odot$ (Bailey et al.\ \cite{Bailey2015})
and a rotation period of $2800\pm200$\,d (Mathys \cite{Mathys2017}).
A temperature of 11\,300\,K is indicative
of a spectral type B8.5p (e.g.\ Eker et al.\ \cite{Eker2020})
rather than the A0p indicated by Renson \& Manfroid (\cite{RensonManfroid2009}).
Hensberge (\cite{Hensberge1993}) already reported
long-term photometric variability with peak-to-peak amplitudes of $\sim0.03$\,mag
and determined a period of the order of 2700\,d.
Bailey et al.\ (\cite{Bailey2015}) estimated that HD\,94660
has completed less than half of its main sequence lifetime.
Its parallax is $8.75\pm0.16$\,mas (Gaia Collaboration et al.\ \cite{Gaia2018}).
Mathys et al.\ (\cite{Mathys1997}) reported variability of the radial velocity indicative
of a spectroscopic binary with an orbital period somewhat longer than two years. 
These variations, with a total range of 35\,km\,s$^{-1}$,
were confirmed by Bailey et al.\ (\cite{Bailey2015}) using UVES (UV-visual echelle spectrograph),
HARPSpol (High Accuracy Radial velocity Planet Searcher polarimeter),
and ESPaDOnS (Echelle SpectroPolarimetric Device for the Observation of Stars) spectra. 
These authors reported that the most likely period of these variations is of the order of 840\,d.
The most recent orbital elements of the system were derived by Mathys (\cite{Mathys2017})
with an orbital period $P_{\rm orb} = 848.96 \pm 0.13$\,d,
an eccentricity $e = 0.4476 \pm 0.0049$,
and a mass function $f(M) = 0.3631 \pm 0.0075$\,${\rm M}_\odot$.
According to the mass function, the secondary must be rather massive ($M > 2\,{\rm M}_\odot$),
yet there is no hint of it in visible spectra.
This led Bailey et al.\ (\cite{Bailey2015}) to the conclusion
that the companion might be a compact object,
such as a neutron star, a black hole, or a pair of white dwarfs.

While about two thirds of the Galactic massive stars appear to be members of close binary systems 
(Sana et al.\ \cite{Sana2012}),
the study of Carrier et al.\ (\cite{Carrier2002})
indicated a scarcity of magnetic Ap/Bp stars in close binaries.
On the other hand, the occurrence of Ap/Bp stars in wide binaries is relatively high. 
In the most recent study of 43~Ap/Bp stars with magnetically resolved lines by Mathys (\cite{Mathys2017}),
22~stars are binaries, with a shortest orbital period $P_{\rm orb}$ of those systems of 27\,d.
The observed dearth of magnetic Ap/Bp stars in close binaries is probably related 
to the origin of the magnetic fields in the Ap/Bp stars.
Recently proposed scenarios suggest
that these stars result from the merging of two lower mass stars or protostars
(Tutukov \& Fedorova \cite{TutukovFedorova2010}; Ferrario et al.\ \cite{Ferrario2009}). 
The merged star would undergo a brief period of strong differential rotation,
which gives rise to large-scale magnetic fields in the radiative envelope.

To better understand the nature of the companion,
we have carried out interferometric observations of HD\,94660 with
PIONIER (Precision Integrated-Optics Near-infrared Imaging ExpeRiment)
on the Very Large Telescope Interferometer (VLTI)
within the framework of our PIONIER Bp multiplicity survey,
looking for wider companions that could have helped an inner binary pair to merge.
Further, we analyzed {\em TESS} (Transiting Exoplanet Survey Satellite) photometric observations
and {\em Chandra} X-ray data.
Finally, we discuss the results of our multiwavelength observations of this system.

\section{Interferometric observations}
\label{sect:observations}

\begin{figure}
\centering
\includegraphics[width=0.43\textwidth]{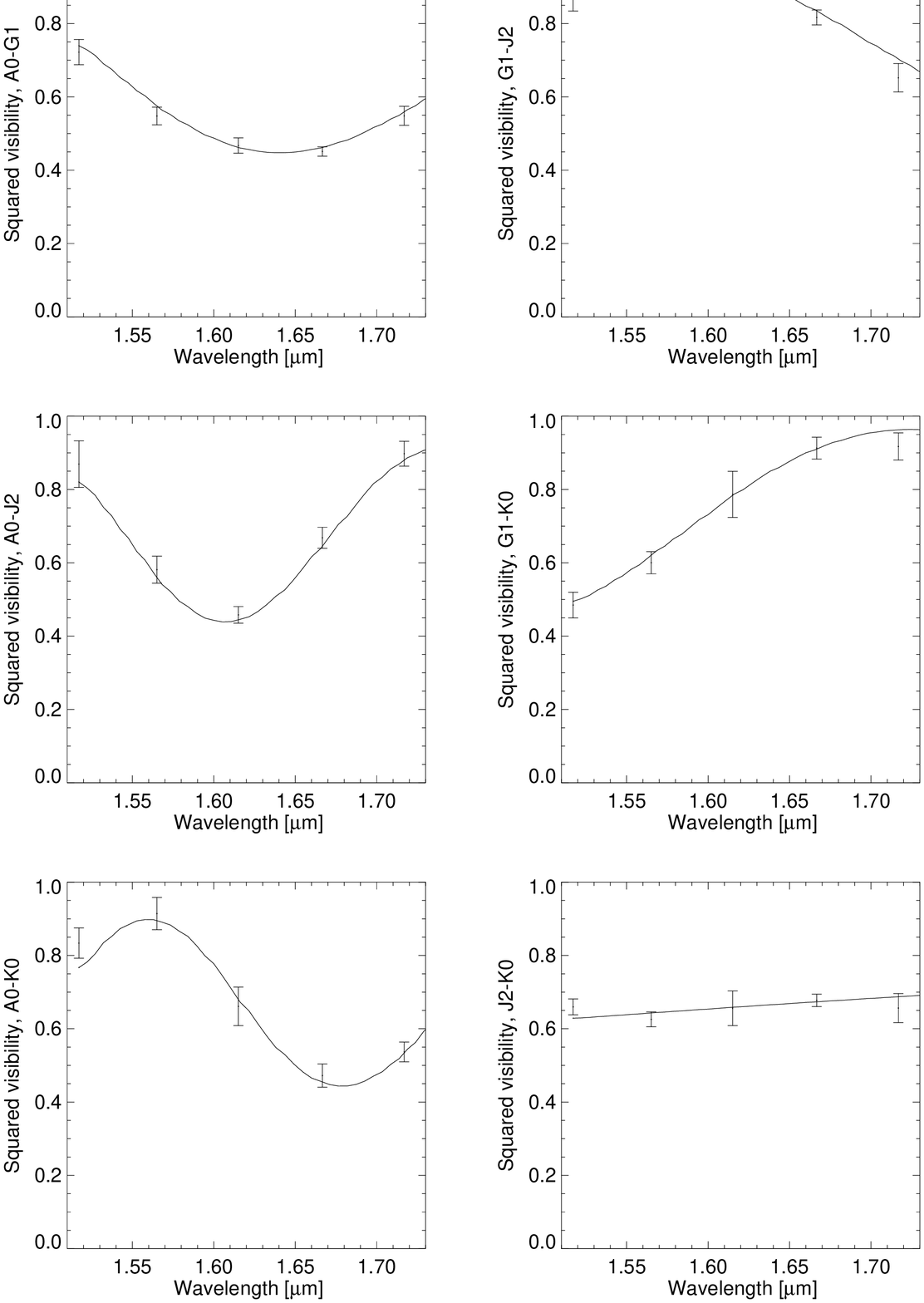}
\includegraphics[width=0.43\textwidth]{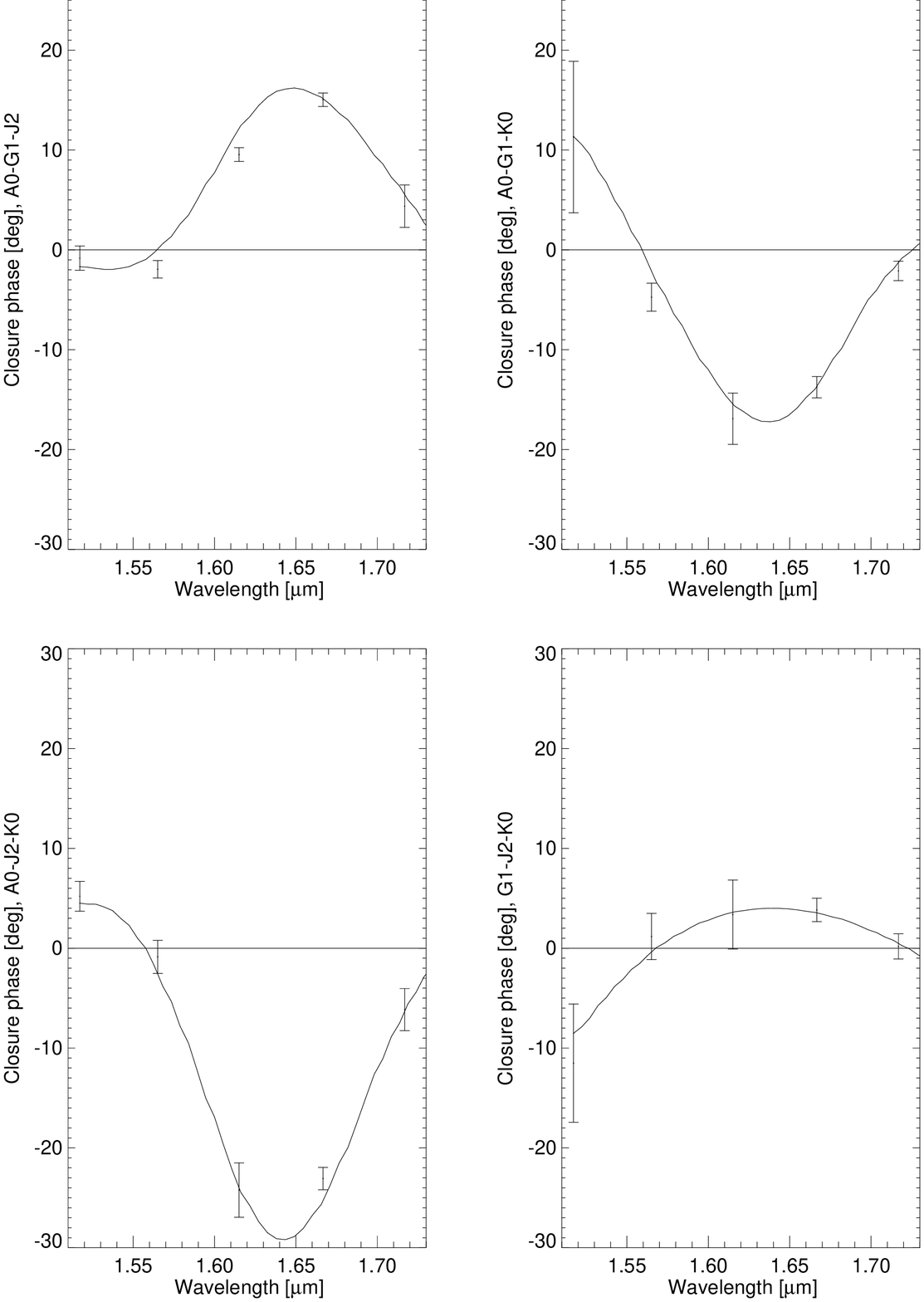}
\caption{
Six visibilities (top) and four closure phases (bottom) of the first PIONIER observations of HD\,94660.
The curves are the binary fit to the data.
}
\label{fig:94_measures}
\end{figure}

We obtained one observation of HD\,94660
with the PIONIER beam combining instrument (Le Bouquin et al.\ \cite{LeBouquin2011})
on the VLTI
(e.g.\ Sch\"oller et al.\ \cite{Schoeller2007})
on 2018 May~18.
PIONIER combines four beams in the H-band,
either from the 8.2\,m Unit Telescopes (UTs), or, as in our case,
from the 1.8\,m Auxiliary Telescopes (ATs),
which leads to visibilities on six different baselines,
as well as four closure phase measurements, simultaneously.
We used PIONIER's low resolution spectroscopic optics
to measure at five different wavelengths within the H-band,
improving the $uv$-coverage.
Alternatively, the instrument can also operate in integrated light
for sensitivity enhancement on faint targets.

All PIONIER data were reduced by us with the {\it pndrs} pipeline
(Le~Bouquin et al.\ \cite{LeBouquin2011}).
It uses the diameters listed in the
JSDC catalog (Chelli et al.\ \cite{Chelli2016}) for the interferometric
calibrator star observed before and after the science target to
compute the transfer function used to calibrate the squared visibility
amplitudes and the closure phases of the science target.
In our case, HD\,95370, with a diameter of 0.52\,mas (Lafrasse et al.\ \cite{Lafrasse2010}),
served as calibrator.
HD\,95370 is not listed in the bad calibrator (BadCal)
list\footnote{www.jmmc.fr/badcal/show.jsp?type=all\&display=simple}.

\begin{figure*}
\centering
\includegraphics[width=0.9\linewidth,angle=0]{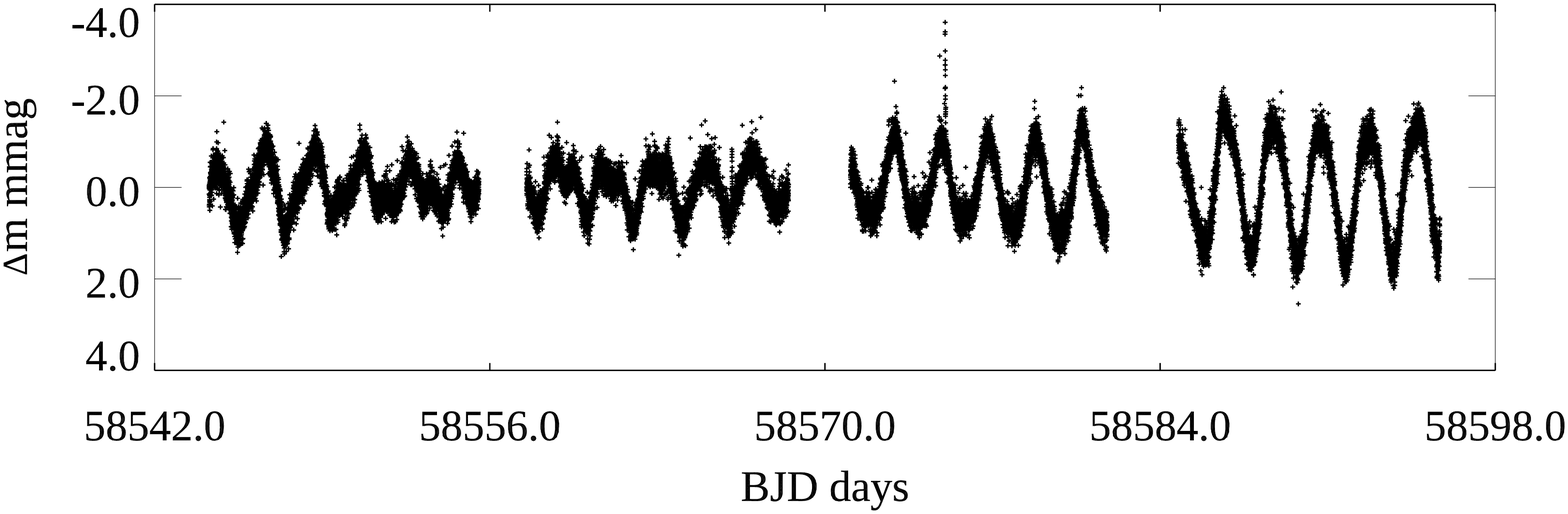}
\includegraphics[width=0.45\linewidth,angle=0]{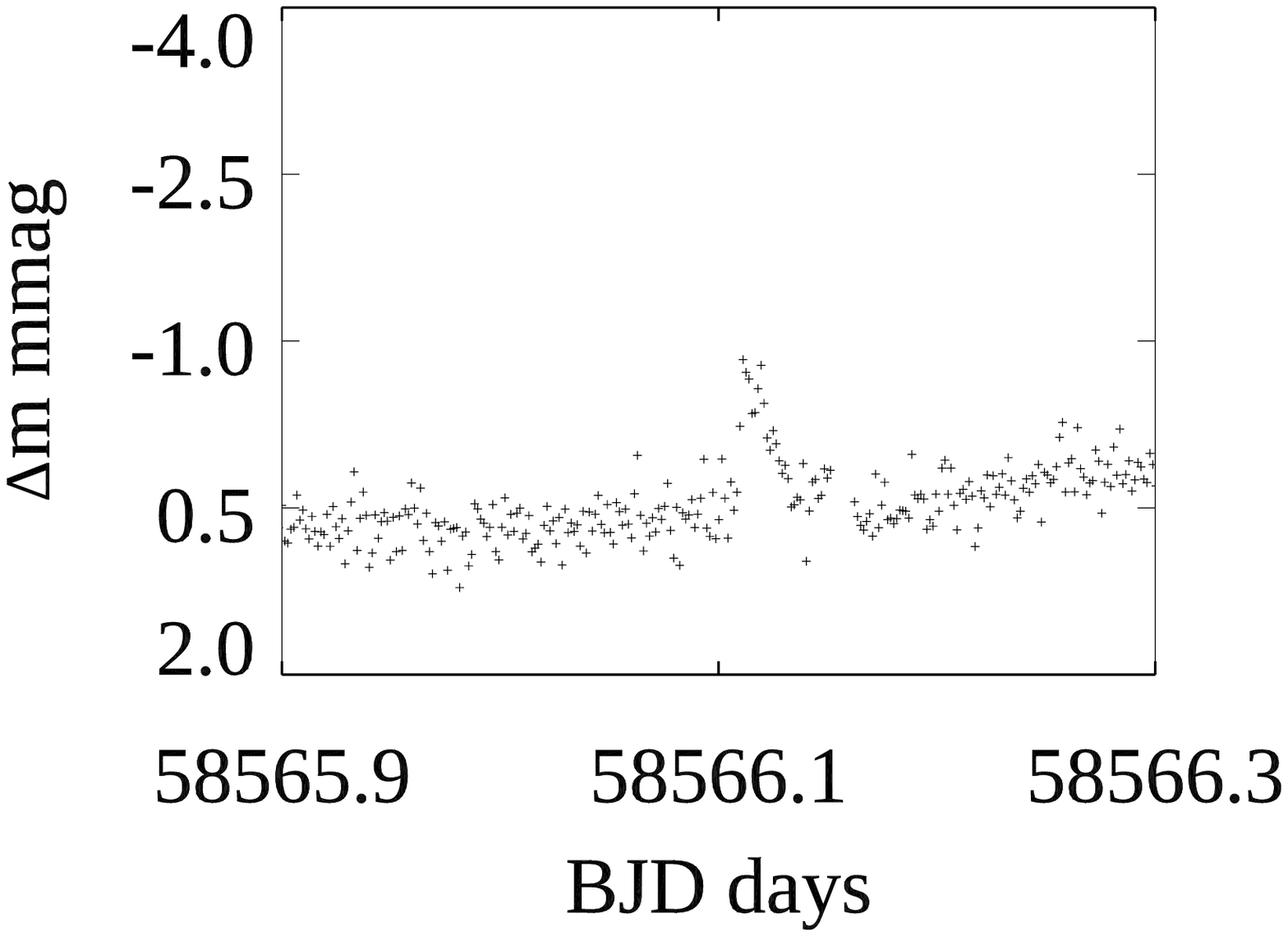}
\includegraphics[width=0.45\linewidth,angle=0]{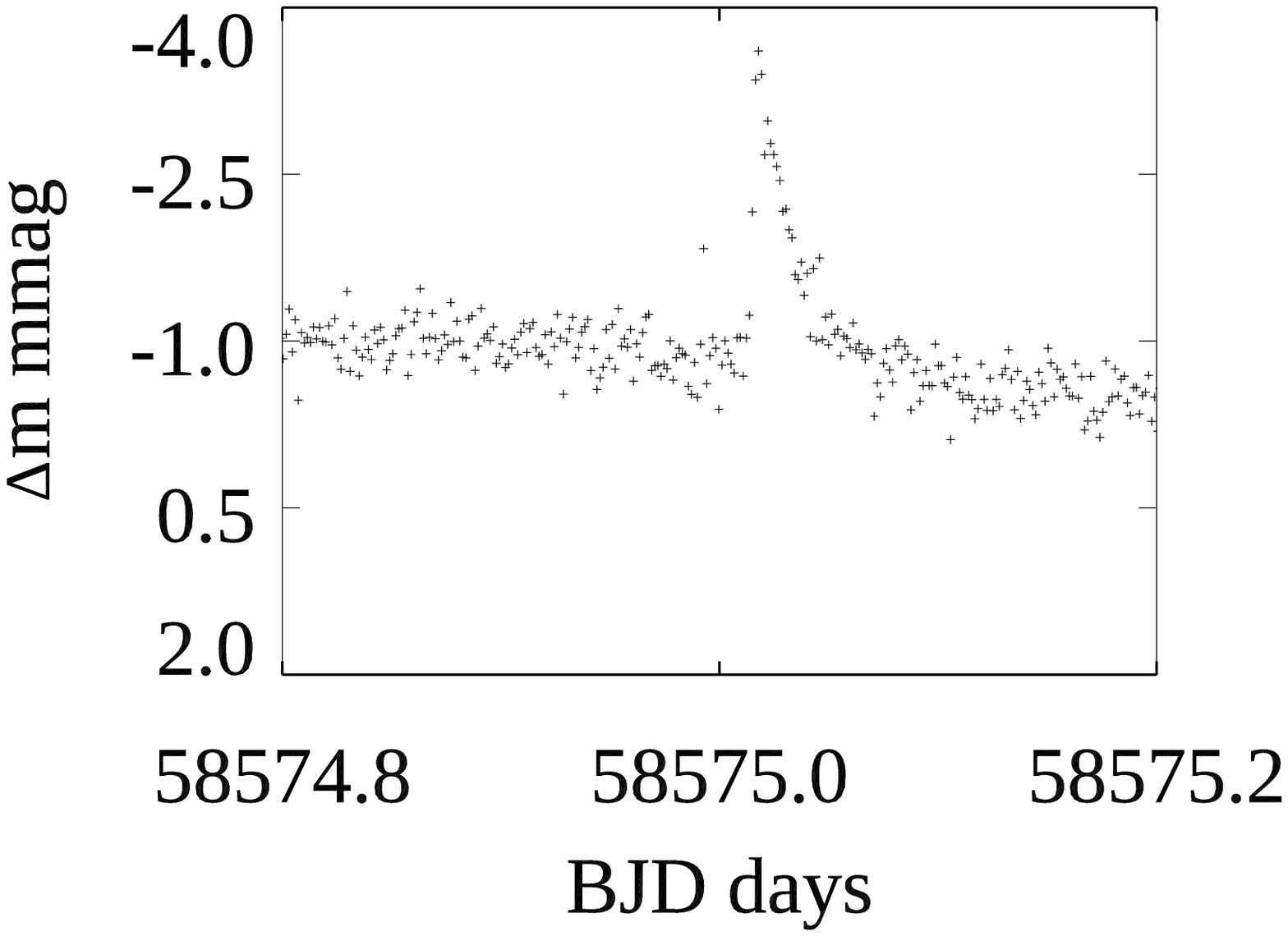}
\caption{
{\sl Top:}
Full Sectors 9 and 10 light curve of TIC\,147622676
showing variations indicative of rotation in a cool, spotted star.
{\sl Bottom:}
Two flares seen after a highpass filter removed much of the low frequency rotational and instrumental variations.
The flares are also indicative of a cool, active star.
The times are bolocentric Julian Date (BJD)$-2400000.0$.
}
\label{fig:lc1}
\end{figure*}

We used the Python tool
CANDID\footnote{github.com/amerand/CANDID}
(Companion Analysis and Non-Detection in Interferometric Data)
developed by Gallenne et al.\ (\cite{Gallenne2015}) to search for a companion in the separation
range up to 40\,mas (limited by bandwidth smearing).
CANDID performs a search on an adaptive grid with
a spacing that will allow it to find the global minimum in $\chi^2$.
For assessing the significance of a companion detection, not only the
reduced $\chi^2$ is considered, but also the number of degrees of freedom.
To support a more stable search with CANDID, we constrained the diameter
of the (primary) star to be equal to its surface brightness diameter
(Mozurkewich et al.\ \cite{Mozurkewich2003}).

We detect a companion to HD\,94660 with an H band magnitude difference of $1.8\pm0.03$
at a separation of $18.72\pm0.02$\,mas and a position angle of $262.74^{\circ}\pm0.05^{\circ}$ (east of north).
The error ellipse, following the CLEAN beam, is $0.07 \times 0.03$\,mas, with a position angle of 6$^{\circ}$.
The binary fit to the data is highly convincing,
with no hints for further components (see Fig.~\ref{fig:94_measures}).

The angular separation corresponds to a projected linear separation of slightly more than 2\,au,
given HD\,94660's parallax of 8.75\,mas (Gaia Collaboration et al.\ \cite{Gaia2018}).
Depending on the total mass of the system
(we use 5\,${\rm M}_\odot$)
and the real linear size of the orbit, this
points to an orbital period of at least 1.4\,yr,
which is of the same magnitude as
the orbital period of 849\,d found by Mathys (\cite{Mathys2017}).
Therefore, we conclude that we indeed have found with PIONIER
the so-far unseen spectroscopic companion.
The H band magnitude difference is not consistent with a compact companion.

\section{{\em TESS} photometric observations}
\label{sect:photometry}

{\em TESS} observations of HD\,94660 (TIC\,147622676) were obtained in Sectors~9 and 10 in 120\,s cadence,
as seen in the top panel of Fig.~\ref{fig:lc1}.
Those observations show significant periodic variations on a time scale of 2.1\,d.
These are variable in both shape and amplitude,
as is characteristic of cool stars showing rotational variation from time-variable spots.

There are two clear instances of flares in the {\em TESS} data.
These are easier to see with the rotational and some instrumental variations filtered from that data,
as shown in the bottom panels of Fig.~\ref{fig:lc1}.
Both flares show typical fast rise and then decay on a time scale of about $30-60$\,min.
The observed amplitudes are only $\sim 1$ and 2\,mmag,
but the intrinsic amplitudes are much larger when taking into account
the brighter primary component of HD\,94660.

If we take an F8V star as an example of a cooler flare star with variable spots,
then the absolute magnitude difference between this star and
the primary B8.5p star is about 4.
Thus, we find the supposed F8V star contributes about 2.5\% of the light of the system.
Scaling up by the inverse of that suggests the flares on the cool star have amplitudes of order $0.05$\,mag,
which is reasonable for such stars.
We have not included any bolometric correction or possible extinction,
since this is a rough estimate to show the scale of the flares. 

Clearly, the Bp star cannot be responsible for the 2.1-d photometric variations;
Ap/Bp stars have stable spots with constant amplitude and shape rotational light curves.
A chance alignment with an unrelated source is also very unlikely,
as no suitable source is known in the sparsely populated field.

\section{X-ray emission}
\label{sect:xrays}

X-ray observations of HD\,94660 were obtained by the {\em Chandra} observatory
with ACIS-I (Advanced CCD Imaging Spectrometer) on 2016 August~20
with an exposure time of 25\,ks (ObsID: 17745).
The {\em Chandra} data processing was done with the CIAO (Chandra Interactive Analysis of Observations)
4.10 software package
and its standard tools were used to produce light curves and spectra.
Source photons were extracted from a circular region with a 5\arcsec{} radius,
the background was taken from a nearby source free region.
The selected energy range is 0.3--10.0\,keV for our source products.

\begin{figure}
\centering
\includegraphics[width=0.45\textwidth]{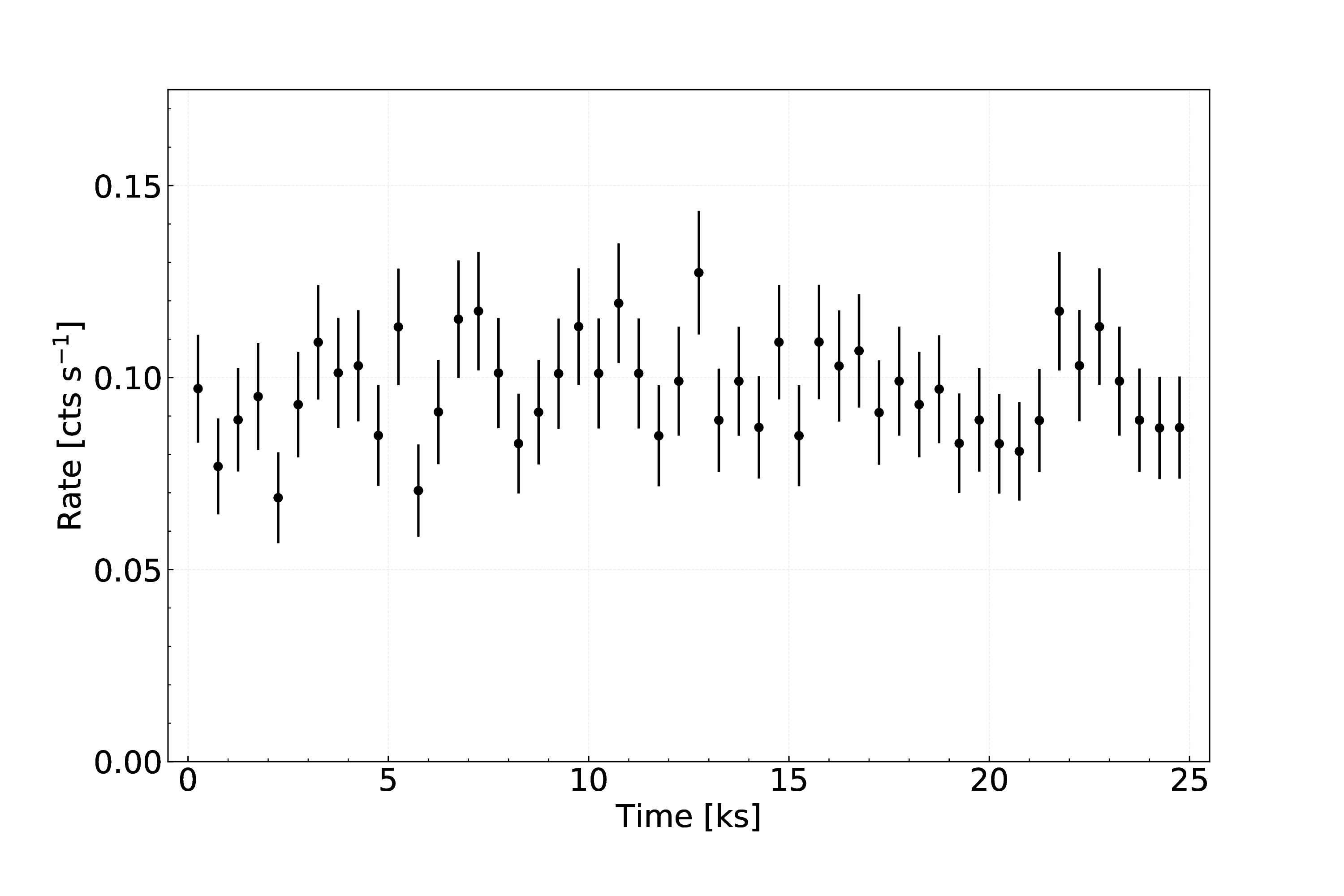}
\caption{
X-ray light curve of HD\,94660 from {\sl Chandra} data, binned to 500\,s intervals.
\label{fig:Xray_lc}
}
\end{figure}

The target shows only low variability in the X-ray light curve,
as presented in Fig.~\ref{fig:Xray_lc}.
Thus, we use all data combined for the spectral analysis,
which was carried out with XSPEC\footnote{heasarc.gsfc.nasa.gov/xanadu/xspec/}
(X-Ray Spectral Fitting Package) V12.9.
We used two-temperature plasma models based on APEC (Smith et al.\ \cite{Smith2001}),
in the following called 2APEC,
as well as an APEC plus powerlaw model, in the following called APEC+POW,
to fit the X-ray spectrum.
The spectrum was rebinned for modeling, so that the errors are 1$\sigma$.
Other models (e.g.\ APEC plus black body radiation) were tested,
but did not yield meaningful results. 

A potential absorption column and the metallicity were treated
as free parameters in the initial modeling.
However, the absorption ($n_{\rm H}$) was found to be consistent with zero
and therefore neglected in the further modeling.
The metallicity is poorly determined and we adopt the best fit value of 0.4~solar,
relative to values given by Grevesse \& Sauval (\cite{GrevesseSauval1998}).
Adding further model components or free parameters like individual element abundances
did not lead to constrained parameters
and did not significantly improve the fit.
We adopted the {\em Gaia} distance of 115~pc.

\begin{table}[t]
\caption{X-ray spectral fit results for HD~94660.}
\label{tab:xray_results}
\begin{center}
\begin{tabular}{lrr}
\hline
\hline\\[-3.1mm]
 \multicolumn{3}{c}{2APEC}\\
\hline\\[-3mm]
kT$_1$ & 0.84\,$^{+ 0.07}_{- 0.03}$  & keV  \\[1mm]
EM$_1$ & 8.52\,$^{+ 0.54}_{- 0.54}$ &  $10^{52}$\,cm$^{-3}$\\[1mm]
kT$_2$ & 2.43\,$^{+ 0.26}_{- 0.22}$  & keV  \\[1mm]
EM$_2$ & 8.20\,$^{+ 0.76}_{- 0.76}$ & $10^{52}$\,cm$^{-3}$\\[1mm]
Abundance & 0.4 & solar\\[1mm]
$\chi^{2}_{\rm red}$ {\tiny(d.o.f.)} & 1.18 (93) & \\[1mm]
\hline\\[-2mm]
$L_{\rm X}$ {\tiny (0.3--10.0\,keV)} & 2.0&  $10^{30}$\,erg\,s$^{-1}$\\[1mm]\hline\\[-2mm]
\multicolumn{3}{c}{APEC+POW }\\\hline\\[-3mm]
kT$_1$ & 0.94\,$^{+ 0.03}_{- 0.03}$  &  keV  \\[1mm]
EM$_1$ & 8.63\,$^{+ 0.83}_{- 0.83}$ &   $10^{52}$\,cm$^{-3}$\\[1mm]
Abundance & 0.4 & solar\\[1mm]
$\alpha$ &  2.39\,$^{+ 0.12}_{- 0.13}$  & [PhoIndex]\\[1mm]
norm {\tiny (at 1\,keV)}&  1.68\,$^{+ 0.28}_{- 0.26}$ & $10^{-4}$ph\,keV$^{-1}$\,cm$^{-2}$ \\[1mm]
$\chi^{2}$ {\tiny(d.o.f.)} & 1.20 (93)& \\[1mm]\hline\\[-2mm]
$L_{\rm X}$ {\tiny (0.3--10.0\,keV)} & 2.3 &  $10^{30}$\,erg\,s$^{-1}$\\[1mm]\hline\\[-2mm]
\end{tabular}
\end{center}
\end{table}

\begin{figure}
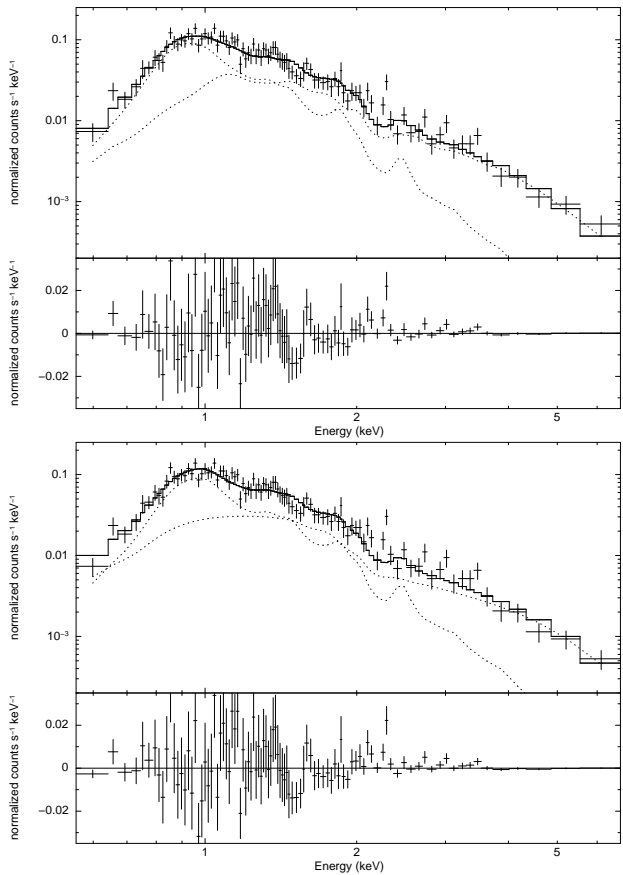

\centering
\includegraphics[angle=270, width=85mm]{KQVel_specb15_2Tres.ps}
\includegraphics[angle=270, width=85mm]{KQVel_specb15_TPowres.ps}
\caption{
X-ray spectrum of HD\,94660 together with our best fit two-temperature thermal model (2APEC, {\sl top})
and thermal plus non-thermal model (APEC+POW, {\sl bottom}).
Each plot shows the data, the model (solid line), the two components (dotted lines),
and the residuals (in the subpanel).
}
\label{fig:specs}
\end{figure}

The 2APEC and APEC+POW models lead to a very similar fit quality, as shown in Fig.~\ref{fig:specs}.
All fit parameters are given in Table~\ref{tab:xray_results}.
Independent of the model details,
we find a flux value of $F_{\rm X} = 1.4 \pm 0.1 \times 10^{-12}$~erg\,cm$^{-2}$\,s$^{-1}$
in the 0.3--10.0\,keV band.
Overall, our analysis agrees within the error bars with the one
presented by Oskinova et al.\ (\cite{Oskinova2020}) from the same data,
making allowance for the solar abundances adopted by these authors.

In the thermal case, the X-ray emission can be well explained by active stars.
Its origin would be coronal, i.e.\ magnetic activity, generated by fast rotating `normal' main-sequence stars.
From the observed X-ray luminosity of $\log L_{\rm X} = 30.3$
and the coronal temperature components of 10\,MK and 25\,MK (average $T_{\rm X} \approx 20$\,MK),
an active binary consisting of late-F to early G-type stars would be a valid candidate.
These spectral types are required, as X-rays from magnetic activity saturate
at about $\log L_{\rm X}/L_{bol} \approx -3$
(see e.g.\ G{\"u}del (\cite{Guedel2004}) for an overview).

Other scenarios that are consistent with the observed X-ray properties
are a Herbig~Ae/Be star (e.g.\ Stelzer et al.\ \cite{Stelzer2009})
or a RS\,CVn system (G{\"u}del \cite{Guedel2004}).
However, the required minimum mass and the observed H band flux ratio
are harder to reconcile with these source classes.

Magnetic Ap/Bp stars are also X-ray emitters via magnetically confined wind shocks
(MCWS; Babel \& Montmerle \cite{BabelMontmerle1997})
and their magnetosphere might even generate non-thermal X-rays via auroral emission
(Leto et al.\ \cite{Leto2017}).
An application of this model to the A0p star CU\,Vir is presented
in Robrade et al.\ (\cite{Robrade2018})
and while its spectral X-ray characteristics are comparable to those of HD\,94660,
its X-ray luminosity is about two orders of magnitudes lower.
Although the magnetosphere of the two stars differ due to the higher magnetic field
and the slower rotation of HD\,94660,
$\log L_{\rm X} > 30$ is above what is typically observed for comparable stars.
Thus, the primary B8.5p star might very well contribute to the observed X-ray emission,
but it is unlikely that it is the sole X-ray source in the system.

\section{Discussion}
\label{sect:disc}

The presence of a compact companion to HD\,94660
is inconsistent with our detection of this
companion with PIONIER on the VLTI with an H band magnitude difference of 1.8.
A compact companion would not be the source of the flares seen by {\em TESS}
and is not required to explain the X-ray emission from HD\,94660.

The {\em Gaia} parallax of HD\,94660 is larger than the one determined by {\em Hipparcos}
(HIgh Precision PARallax COllecting Satellite)
($8.75\pm0.16$\,mas vs.\ $6.67\pm0.80$\,mas; ESA \cite{ESA1997}).
Bailey et al.\ (\cite{Bailey2015}) used the {\em Hipparcos} parallax to calculate
a distance of 150\,pc to HD\,94660.
Using the {\em Gaia} parallax, this would lead to 115\,pc,
the luminosity would drop down from 105\,${\rm L}_\odot$ to 62\,${\rm L}_\odot$,
and the mass from 3.0\,${\rm M}_\odot$ to 2.6\,${\rm M}_\odot$.
The lower mass does however not agree with the temperature of 11\,300\,K,
determined by Bailey et al.\ (\cite{Bailey2015}),
which might be a sign for current {\em Gaia} parallaxes not being reliable
for bright stars as well as for binary objects.
A full interferometric and spectroscopic orbital solution of HD\,94660
will resolve this discrepancy.
But even if we use this lower primary mass of 2.6\,${\rm M}_\odot$ in the system's mass function,
the mass of the companion is still above 2.0\,${\rm M}_\odot$.

The first interferometric observation, carried out in the framework of our PIONIER Bp star multiplicity survey,
showed that HD\,94660 is a binary with a separation of 18.72\,mas.
The binary fit to the data shows no hints of further companions.
The angular separation found corresponds to a projected linear separation of slightly more than 2\,au.
This points to an orbital period of at least 1.5\,yr,
which is of the same magnitude as
the orbital period of 849\,d found by Mathys (\cite{Mathys2017}).
Therefore, we conclude that we indeed have found with PIONIER
the so-far unseen spectroscopic companion.
We checked once more the spectra from UVES and HARPSpol used by Bailey et al.\ (\cite{Bailey2015})
and could not find any trace of a companion.

HD\,94660 was also observed by {\sl TESS}.
For extremely slowly rotating Ap/Bp stars, {\sl TESS} observations generally show a flat light curve
with no signal above the noise level in the Fourier transform.
However, for HD\,94660, {\sl TESS} observed variations of significant amplitude (several mmag peak-to-peak),
which are cyclic on a timescale of the order of 2.1\,d,
but not periodic -- there are considerable differences from cycle to cycle --
with occasional flare-like events superimposed.
Clearly, the primary Bp star cannot be responsible for the photometric variations.
A chance alignment with an unrelated source is very unlikely,
as no suitable source is known in the sparsely populated field.
The variations are very reminiscent of those of active late-type stars or pre-main sequence stars.

Our X-ray observation with {\em Chandra} shows slight flux variations
over the total exposure of about 7\,h, but no major flaring.
The spectral properties are compatible with magnetic activity in the companion
and/or a modified magnetically confined wind shock model.

If we try to reconcile all information that we have gathered in our study of HD\,94660,
we have to keep the following in mind:
a)
the H band magnitude difference of 1.8 between primary and secondary,
b)
that the companion is not seen in spectra taken at visible wavelengths,
c)
the companion mass of at least 2\,${\rm M}_\odot$,
d)
that we need a source of photometric variability with a 2.1\,d period in the system,
and e)
that we need a source for flares in the system.
The primary Bp star is not the source of either the photometric variability or the flares.

If we assume that the secondary is a single main-sequence star,
then the mass of at least 2\,${\rm M}_\odot$ would translate into a spectral type earlier than A5V.
The magnitude difference in the visible between primary and secondary would then be
on the order of 1, which is in conflict with the companion not having been seen in the visible spectra.
On the other hand, a main sequence single star that obeys the $\Delta m_H = 1.8$
would correspond to an F5V star,
which can be ruled out too, as its mass is too low, around 1.5\,${\rm M}_\odot$.
A 2\,${\rm M}_\odot$ giant would again be too bright in the visible and
the evolutionary tracks of both components would be difficult to align.
A pre-main sequence Herbig Ae/Be star would fit the mass,
but should be of brightness similar to the primary in the visible.
Using the VizieR Photometry viewer\footnote{vizier.unistra.fr/vizier/sed/},
we do not see any infrared excess for HD\,94660, making this scenario quite unlikely.
Also, while we expect photometric variations from a Herbig Ae/Be star,
mainly due to the accretion process, they would rather be stochastic than quasi-periodic.

If we go toward a higher order multiple system, then two fast rotating,
young and active F8V stars in a close binary
would together have a mass higher than 2\,${\rm M}_\odot$.
Schmidt-Kaler (\cite{SchmidtKaler1982}) gives for a B8V star an absolute visual magnitude of $-$0.25
and for a B9V star of +0.2.
We adopt for our B8.5p star an absolute visual magnitude of 0.0.
An F8V star has according to Schmidt-Kaler an absolute visual magnitude of +4.0.
Thus, the magnitude difference between one F8 star
and the late B primary in the visible would be around 4.
Given the high quality of some of the spectra of HD\,94660 that have been analyzed
both by Bailey et al.\ (\cite{Bailey2015}) and by us,
one can very conservatively assume that any line of the secondary
that has a depth greater than 1\% of the continuum should have been detected.
With a flux ratio of 40 between the Bp star and one of the two F8V companions,
the corresponding upper limit on the intrinsic depth of a line
in the spectrum of the latter is 40\% of the continuum.
It is not easy to make a quantitative estimate,
but it should not require such a F8V star to have a very large $v\,\sin\,i$
for even its strongest lines to have intrinsic depths of less than 40\%
(see e.g.\ Fig.~6 of Slettebak et al.\ (\cite{Slettebak1975}) for an F5V template spectrum).
We also have to note that it is quite common that flaring stars are found in binaries.
Using the colors provided by Ducati et al.\ (\cite{Ducati2001}),
$(V-H)_0 = -0.40$ for a B8.5V star and $(V-H)_0 = 1.06$ for an F8V star,
we calculate a magnitude difference in the H band between a B8.5V star
and a single F8V star of 2.55 and between a B8.5V star and two F8V stars
of 1.8, matching the magnitude difference measured by us with PIONIER.

In conclusion,
a fast rotating and active F8V+F8V close binary as the companions of the Bp component of the HD\,94660 system
is plausible and compatible with the available observational constraints.
Of course, combinations of two stars of similar spectral types are also able to explain
the observational constraints.

To investigate the nature of the infrared, optically invisible, companion of HD\,94660,
additional interferometric observations are necessary.
These observations, together with the spectroscopic solution,
will allow the determination of the full orbit, further constrain the nature of both components,
and fix the distance toward HD\,94660.
The characterization of the companion will also help to understand the role
of binarity for the origin of magnetism in stars with radiative envelopes.

\begin{acknowledgements}

Based on observations obtained at the European Southern Observatory, Paranal, Chile
(ESO programme
No.~0101.D-0399(B)).
This paper includes data collected by the {\em TESS} mission.
Funding for the {\em TESS} mission is provided by the NASA Explorer Program.
The scientific results reported in this article are based in part 
on observations made by the {\rm Chandra} X-ray Observatory under ObsID 17745 (PI~Hubrig).
We would like to thank the anonymous referee for useful suggestions to the manuscript
and J.~Federico~Gonz\'alez for
fruitful discussions on the nature of the companion.
This research has made use of the Jean-Marie Mariotti Center \texttt{SearchCal}
service\footnote{Available at http://www.jmmc.fr/searchcal}
and the JSDC catalogue\footnote{Available at http://www.jmmc.fr/catalogue\_jsdc.htm}
co-developed by FIZEAU and LAOG/IPAG,
and of the CDS Astronomical Databases SIMBAD and VIZIER.

\end{acknowledgements}

\end{document}